\documentstyle[sprocl,epsfig]{article}

\def\beq{\begin{equation}}
\def\eeq{\end{equation}}
\def\bea{\begin{eqnarray}}
\def\eea{\end{eqnarray}}

\def\as{\alpha_s}

\def\sigmahat{\hat{\sigma}}

\def\Dzero{D$\emptyset$}

\def\GeV{{\rm GeV}}

\def\cG{{\cal G}}

\def\lapprox{\lower .7ex\hbox{$\;\stackrel{\textstyle <}{\sim}\;$}}
\def\gapprox{\lower .7ex\hbox{$\;\stackrel{\textstyle >}{\sim}\;$}}
\def\xbj{x_{\rm Bj}}
\def\qbar{\bar{q}}
\begin{document}
\begin{flushright}                                                                     
DTP/98/24  \\                                                                     
UR-1526\\ 
ER/40685/915\\ 
April 1998 \\
{}~~~~~~~~~~~~                                                                    
\end{flushright}                                                                     

\title{A BFKL MONTE CARLO APPROACH TO JET PRODUCTION AT HADRON--HADRON
AND LEPTON--HADRON COLLIDERS\footnote{Presented by W.J.~Stirling
at the 6th International Workshop on Deep Inelastic Scattering and
 QCD (DIS98), Brussels, 4-8 April 1998.}}

\author{L. H. ORR}

\address{Department of Physics and Astronomy, University of Rochester,\\
Rochester, NY~14627-0171, USA\\E-mail: orr@urhep.pas.rochester.edu} 

\author{W. J. STIRLING}

\address{Departments of Mathematical Sciences and Physics, 
University of Durham,\\
Durham DH1~3LE, England\\E-mail: w.j.stirling@durham.ac.uk}

\maketitle\abstracts{The production of a pair of jets with
large rapidity separation in hadron--hadron collisions, and of forward
jets in deep inelastic scattering, can in principle be used
to test the predictions of the BFKL equation. However in practice
kinematic constraints lead to a strong suppression of BFKL effects
for these processes. This is illustrated  using a BFKL Monte
Carlo approach.}

Perturbative QCD at fixed order  in the
strong coupling constant $\as$ provides 
sufficient predictive power for a wide variety of high energy phenomena.
However in some  regions of phase space, large logarithms can  
multiply the coupling,  
 spoiling the good behaviour of fixed--order expansions.  In certain cases these
 large logarithms can be resummed, as in the 
Balitsky, Fadin, Kuraev and Lipatov (BFKL)  equation \cite{bfkl}, and 
the predictive power of the theory is then in principle restored.  

\begin{figure}[t]
\begin{center}
\mbox{\epsfig{figure=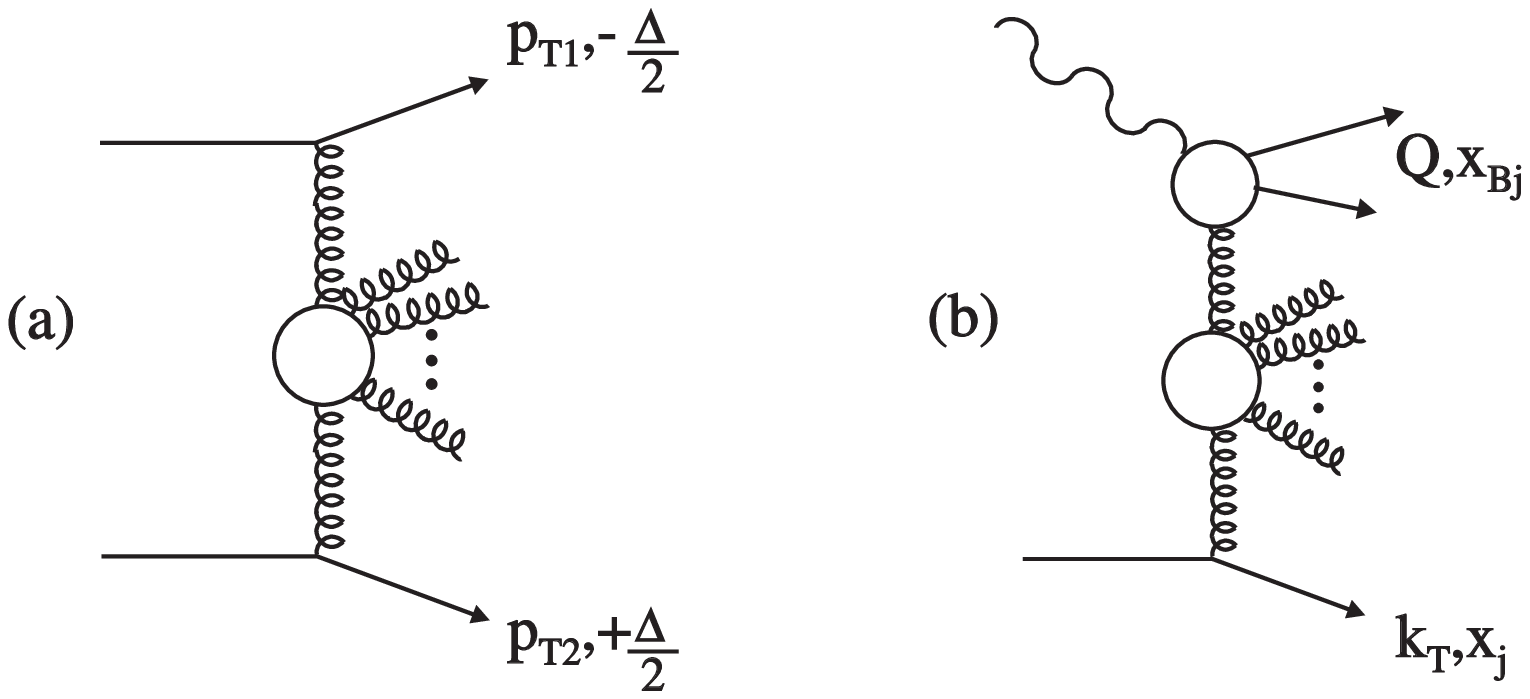,width=10.0cm}}
\caption{Schematic representation of (a) dijet production with large
rapidity separation $\Delta$ in hadron--hadron collisions, and (b) forward
jet production in deep inelastic scattering. Note that the solid lines 
at each end of the gluon ladder in (a) and at the bottom 
of the ladder in (b) represent either quarks or
gluons, according to the effective subprocess approximation.}
\label{fig:ladders}
\end{center}
\end{figure}
At  hadron colliders the BFKL equation applies to
dijet production when the rapidity separation $\Delta$ of the two jets
(with $p_{T1} \sim p_{T2} \sim p_T$)
is large. The emission of (real and virtual)
gluons in the rapidity
interval between the two leading jets (see Fig.~\ref{fig:ladders}a)
generates large logarithmic
contributions $(\as\ln(\hat s/p_T^2))^n \sim (\as\Delta)^n$
 which when resummed  \cite{muenav} give a dijet
subprocess cross
section which increases with $\Delta$,
$\sigmahat_{jj}\sim \exp(\lambda\Delta)$, with $\lambda = 12 \ln 2
\as/\pi\approx 0.5$, in contrast to
the $\sigmahat_{jj} \to \ $constant behaviour expected at lowest order.
In practice 
such behaviour can be difficult to observe because $\sigmahat$
 gets folded in with parton distribution functions (pdfs), which 
decrease with $\Delta$ much more rapidly than the subprocess 
cross section increases:
\beq
\sigma_{jj}\; \sim \; q(x_1)\;  q (x_2)\; \exp(\lambda\Delta)
\eeq
with $x_{1,2} \sim p_T/\sqrt{s} \exp(\Delta)$.
The challenge is to find measurable quantities
in dijet production that are insensitive to the pdfs, but that retain the
distinctive behaviour characteristic of BFKL resummation.
One  possibility \cite{many,crs,os} is
 the azimuthal decorrelation of the two jets:  the multiple
emission of soft gluons between the leading jets predicted by BFKL leads
to a stronger decorrelation than does fixed--order QCD, and the prediction is
relatively insensitive to the pdfs.
Another possibility is to look for the increase in
$\sigmahat$ with $\Delta$ by considering different collider energies
\cite{muenav}.
The idea is to choose $\Delta$'s that correspond to the same parton
momentum fractions at different energies so that the pdf dependence is
the same for both, thereby allowing the $\Delta$ dependence of
$\sigmahat$ to be extracted. This has been studied recently in Ref.~\cite{os2},
see also \cite{ANNA}.

Exactly the same (BFKL) physics applies to the production of
 `forward' $k_T^2 \sim Q^2$ jets in deep inelastic scattering \cite{FJ},
where the large rapidity separation is between the current and forward
jets, see Fig.~\ref{fig:ladders}b.

However, a key issue is whether the kinematic regions 
currently accessible at the Tevatron and at HERA allow the 
asymptotic BFKL conditions to be fulfilled. In particular,
the asymptotic $ \exp(\lambda\Delta) $ behaviour is obtained under the
assumption that there is no kinematic penalty for emitting arbitrary
large numbers of gluons with $k_{Ti} \sim p_T$. In practice,
there {\it is} an overall constraint from $\sqrt{\hat s} < \sqrt{s}$ and even
before this is reached there is a strong suppression of
large subprocess energies from the pdfs. To investigate these effects,
we have constructed \cite{os} a BFKL Monte Carlo programme (BFKL--MC)
which reproduces the analytic predictions by explicitly taking
into account the  emission
of real and virtual gluons to all orders.\footnote{The technical details
can be found in Ref.~\cite{os}. A similar technique has been used
in Ref.~\cite{crs}.} With such an event generator the
effects of kinematic constraints and
 experimental cuts can readily be taken into account.

\begin{figure}[t]
\begin{center}
\mbox{\epsfig{figure=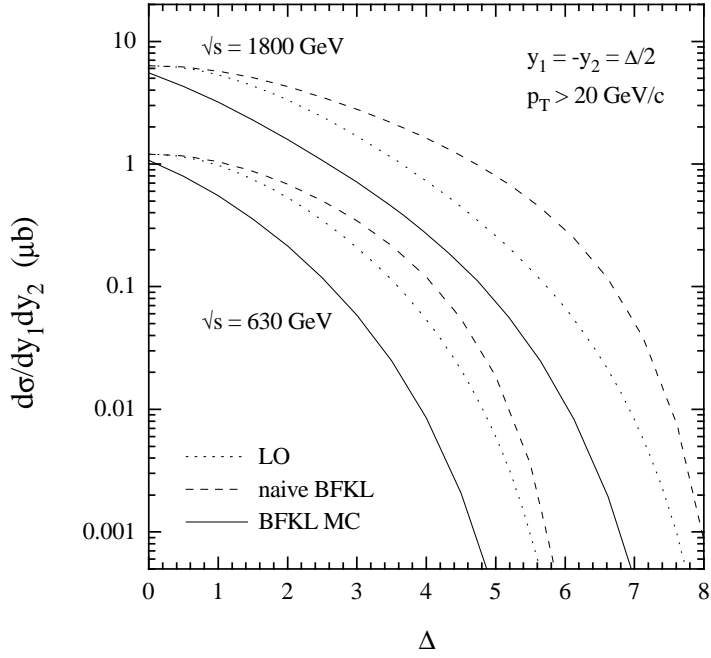,height=10cm}}
\vspace{-0.5cm}
\caption{The dependence of the BFKL and asymptotic QCD leading--order
dijet cross sections  
on the dijet rapidity separation. The three curves at each collider energy
 use: (i) `improved' BFKL--MC (solid lines), (ii) `naive' BFKL (dashed lines),
  and (iii) the asymptotic ($\Delta \gg 1$) form
 of QCD leading order (dotted lines).}
\label{fig:sigmabfkl}
\end{center}
\end{figure}

Fig.~\ref{fig:sigmabfkl} shows the dijet cross section as a function of
$\Delta$ 
for the naive BFKL and improved BFKL--MC
cases at $\sqrt{s} = 630, 1800\;\GeV$, with
$p_{T1}, p_{T2} > 20\; \GeV$.
Asymptotic QCD LO is also shown for reference.
The naive BFKL
cross section (dashed curve) is always largest, because it includes the 
analytic 
subprocess cross section $\hat\sigma_{jj}\sim \exp(\lambda\Delta)$,
which corresponds to the emission
of any number of gluons with arbitrarily large energies.  The prediction
falls off rather than increases because $\hat{\sigma}$ is multiplied
by the pdfs, but even those incorporate only  lowest order
kinematics in this case.
When exact kinematics for entire events are included in both the 
subprocess cross section and the pdfs, as in the BFKL--MC
(solid curve), there is a dramatic suppression of
the total cross section.\footnote{The running of $\alpha_s$,
which we include, 
also contributes to the suppression, but it has a much smaller
effect than kinematics.}  In fact the suppression is so strong that 
it drives the BFKL--MC cross section {\it below} that for asymptotic QCD LO.
The reason is due to simple kinematics: 
the QCD LO cross 
section contains only two jets, but the BFKL--MC cross section also includes
additional jets, each of which increases the subprocess centre--of--mass energy
and elicits a corresponding price in parton densities.
In the naive BFKL calculation, the contribution to the subprocess energy
from additional jets is ignored and their net effect is to combine
with  the virtual gluons to increase the subprocess cross section.

Finally we consider forward jet production in deep inelastic scattering
at HERA, Fig.~\ref{fig:ladders}b. It is relatively straightforward 
to adapt the dijet formalism
to calculate the cross section for the production of a forward jet with a given
$k_T$ and longitudinal momentum fraction $x_j \gg \xbj$. In fact there
is a direct correspondence between the variables: $p_{T2} \leftrightarrow 
k_T$ and $\Delta \leftrightarrow \ln(x_j/\xbj)$. In the DIS case the variable
$p_{T1}$ corresponds to the transverse momentum of the $q \bar q$ pair
in the upper `quark box' part of the diagram. In practice this variable
is integrated with the off--shell $\gamma^* g^* \to q \bar q$ amplitude
such that $p_{T1}^2 \sim Q^2$. As a result, it is appropriate to consider
values of $k_T^2$ of the same order, and to consider the (formal) 
kinematic limit $x_j/\xbj \to \infty$, $Q^2$ fixed. In this limit we
obtain the `naive BFKL' prediction $\hat \sigma_{\rm jet} \sim (x_j/\xbj)^\lambda$,
the analogue of $\hat\sigma_{jj} \sim \exp(\lambda\Delta)$.

\begin{figure}[ht]
\begin{center}
\mbox{\epsfig{figure=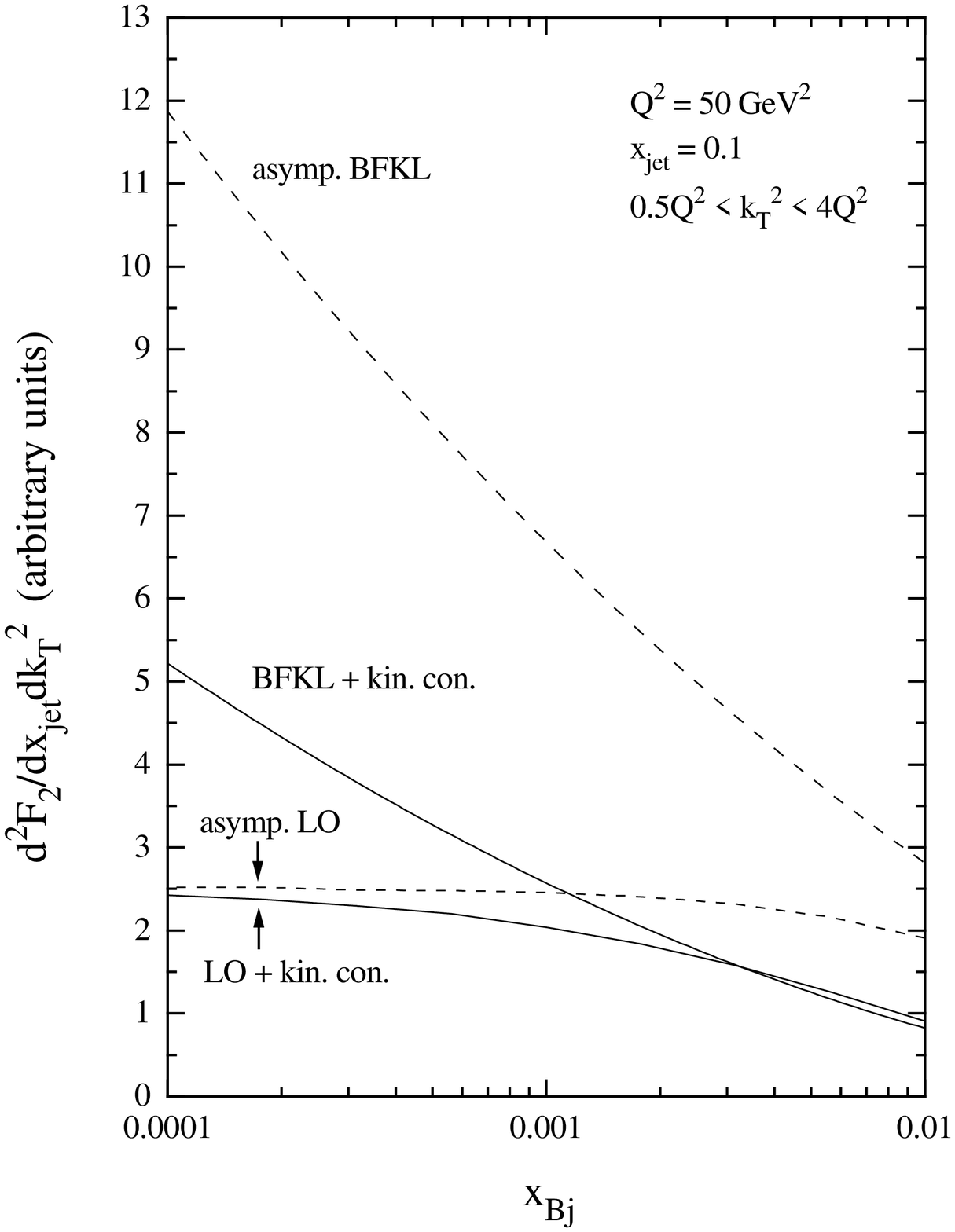,width=12.0cm}}
\vspace{-0.5cm}
\caption{Differential structure function for forward jet production in
$ep$ collisions at HERA. The curves are described in the text.}
\label{fig:djhera}
\end{center}
\end{figure}
Fig.~\ref{fig:djhera} shows the differential structure function
$\partial^2 F_2/\partial x_j\partial k_T^2$ as a function of $\xbj$ at HERA,
with 
\beq
x_j = 0.1, \qquad Q^2 = 50\; \GeV^2,\qquad Q^2/2 < k_T^2 < 4 Q^2.
\eeq
The lower dashed curve is the leading--order prediction from the 
process $\gamma^* \cG \to q \qbar \cG$, with $\cG = g,q$, with no 
overall energy--momentum constraints. This is the analogue of the
$\hat\sigma_{jj} \to\ $constant prediction for dijet production. Note that
here the parton distribution function at the lower end of the ladder
is evaluated at $x = x_j$, independent of $\xbj$. In practice, when 
$\xbj$ is not small we have $x > x_j$ and the cross section is suppressed,
as indicated by the lower solid curve in Fig.~\ref{fig:djhera}. The upper
dashed curve is the asymptotic BFKL prediction with the characteristic
$(x_j/\xbj)^\lambda$ behaviour. Finally the upper solid line is the
prediction of the full BFKL Monte Carlo, including kinematic constraints
and pdf dependence. There is evidently a significant suppression of the 
cross section. We emphasise that Fig.~\ref{fig:djhera} corresponds to
`illustrative' cuts and should not be directly compared to the experimental
data. Nevertheless, the BFKL--MC predictions do appear to
follow the general trend of the H1 and ZEUS measurements \cite{HERA}.
A more complete study, including realistic experimental cuts and an
assessment of the uncertainty in the theoretical predictions, is under way
and will be reported elsewhere \cite{os3}.

\end{document}